\documentclass[a4paper, 12 pt]{article}
\usepackage{amsfonts}
\usepackage{amsthm}
\usepackage{amscd}
\usepackage{ulem}
\usepackage{cancel}
\usepackage{amssymb}
\usepackage{graphicx}
\usepackage{color}
\usepackage{graphics}
\usepackage{setspace}

\newcommand{\third}{{\textstyle \frac{1}{3}}}

\theoremstyle{definition}

\theoremstyle{remark}

\newcommand{\dst} {\displaystyle}

\newcommand{\lb} {\left(}
\newcommand{\rb} {\right)}

\newcommand{\ls} {\left[}
\newcommand{\rs} {\right]}

\newcommand{\al} {\alpha}

\newcommand{\bga}{\begin{array}{l}}
\newcommand{\ena}{\end{array}}
\newcommand{\bge}{\begin{equation}}
\newcommand{\bgea}{\begin{equation} \begin{array}{l} }
\newcommand{\ene}{\end{equation}}
\newcommand{\enea}{ \end{array} \end{equation}}

\newcommand{\text}{ }
\begin{document}
\fontsize{10pt}{11pt}\selectfont
\title{Reactive self-heating model of aluminum spherical nanoparticles}
\author{Karen S Martirosyan$^{\dag}$\thanks{karen.martirosyan@utb.edu}, Maxim Zyskin$^{\ddag}$, \\\\
{$^{\dag}$Department of Physics and Astronomy,} 
{University of Texas, Brownsville,}\\
{80 Fort Brown, Brownsville, TX, USA, 78520;}\\
{$^{\ddag}$Rutgers University, 126 Frelinghuysen Road, Piscataway, NJ 08854-8019.}\\
\\}

\date{ }

\maketitle
\begin{doublespace}

\section*{Abstract} 
Aluminum-oxygen reaction is important in many highly energetic,  high pressure generating systems. Recent experiments with nanostructured thermites suggest that oxidation of aluminum nanoparticles occurs in a few microseconds. Such rapid reaction cannot be explained by a conventional diffusion-based mechanism. We present a rapid oxidation model of a spherical aluminum nanoparticle, using Cabrera-Mott moving boundary mechanism, and taking self-heating into account. In our model, electric potential solves the nonlinear Poisson equation. In contrast with the Coulomb potential, a "double-layer" type solution for the potential and self-heating leads to enhanced oxidation rates. At maximal reaction temperature of 2000 $^o$C, our model predicts overall oxidation time scale in microseconds range, in agreement with experimental evidence.
\newpage

Many previous studies have been directed to understand the mechanism and kinetics of aluminum particle oxidation.
\cite{drez} -\cite{11}.
Aluminum oxidation exhibits high enthalpy and has been extensively used for propulsion, pyrotechnics and explosion reactions \cite{1} \cite{2}.  Nanoenergetic materials (NM) based on aluminum thermites may store two times more energy per volume than conventional monomolecular energetic materials \cite{3}. The size reduction of reactant powders such as aluminum from micro- to nano-size increases the reaction front propagation velocity in some systems by two to three orders of magnitude
\cite{3} \cite{4}. The development of novel NM, their design, synthesis and fabrication procedures are critical for national security and it was recognized to be significant to support advanced weapons platforms. Among numerous thermodynamically feasible Metastable Intermolecular Composites mixtures the most widely investigated are  \ Al/Fe$_2$O$_3$, \ Al/MoO$_3$,    Al/WO$_3$, Al/CuO, Al/Bi$_2$O$_3$ and Al/I$_2$O$_5$ nano systems \cite{3}-\cite{9+}. The main distinguishing features of these reactive systems are their significant enthalpy release and tunable rate of energy discharge, which gives rise to a wide range of combustion rates, energy release and ignition sensitivity.

There are several advantages of using  Al/Bi$_2$O$_3$ and Al/I$_2$O$_5$ nanocomposites: (i) reduced ignition  and reaction times; (ii) superior heat transfer rates; (iii) tunability of novel energetic fuel/propellants with desirable physical properties; (iv) enhanced density impulse; (v) incorporating nanoenergetic materials into the MEMs and NEMs systems \cite{3}. Our recent experiments suggest that oxidation of nanoparticles of aluminum with
Bi$_2$O$_3$ and I$_2$O$_5$ occurs in a few microseconds
\cite{7}-\cite{9}.  Rapid reaction in the nanostructured thermites cannot be explained by a conventional mechanism based on the diffusion of Al and O atoms in oxides.

In this report, we present a rapid oxidation model of spherical aluminum nanoparticles surrounded by oxygen, using Cabrera-Mott oxidation model
\cite{mott}-\cite{11} with a self-consistent potential, and taking self-heating into account.
Using nonlinear self-consistent potential in Cabrera-Mott model, as opposed to the Coulomb potential, is more accurate, and yields higher oxidation rates. For nanosized particles, nonlinear model gives "double-layer" type solution for the potential, since potential changes rapidly near the metal-oxide interface.
As a result, the electric field at the metal- oxide interface, determining the oxidation rate, is less sensitive to oxide thickness, and stays nearly constant at intermediate stages of oxidation (while for the Coulomb potential, electric field decreases, and the oxidation rate quickly drops).
This nonlinear effect for the potential, combined with  self-heating, leads to  rapid  temperature increase, resulting in a dramatic (orders of magnitude) increase of oxidation rates for nano-sized particles.
Our model can predict oxidation times in microseconds range, in a good agreement with  experimental data \cite{3}.

To estimate the reaction times, we assume that  aluminum sphere (radius  $25$ nm,  with thin oxide layer of $3$ nm)  is surrounded by   oxygen. The sphere is rapidly heated to ignition  temperature $T_0,$ sufficient to initiate  oxidation reaction, further boosted  by self-heating as a result of oxidation. We assume the spherical symmetry of the problem.

In the Cabrera-Mott model of metal oxidation \cite{drez}-\cite{15}, \cite{mott}  aluminum ions are helped to escape aluminum boundary (overcoming ionization potential $W$)
by a self-consistent electric   potential $V.$ The electric field is induced in the oxide layer by imbalance between excess positive aluminum ions and electrons in the oxide layer, due  to difference in chemical potentials of the metal and oxidizer.
Excess concentration of electrons and ions in the oxide layer  depends on the electric field potential $V $  via appropriate Gibbs factors, leading to a self-consistent version of Poisson equation for $V,$   which in the spherically-symmetric case is given by  
\bgea
\dst \nabla^2 V \equiv  \frac{1}{r^2}\frac{d}{dr} \lb r^2 \frac{d}{dr} V \rb = 8 \pi k_0 \mathbf{e}\  N \sinh\lb \frac{ \mathbf{e}  V}{k_b T } \rb ,  \quad
r_1 \leq r \leq r_2 ,
\\ \\
V \lb r_1 \rb = V_0,   
\\
V \lb r_2 \rb = 0 .
\label{MottPot}
\enea
Here $r_1$ is the metal particle radius, $(r_2-r_1)$ is oxide layer thickness,    $k_0 = 8.99 \ 10^9 \ \frac{ \mbox{N m}^2 }{\mbox{C}^2},$ $\mathbf{e} = 1.60 \ 10^{-19} \mbox{C},$
$N = \lb n_e n_i \rb^{1/2},$ and  $n_i$ and $n_e$ are concentrations of respectively aluminum ions   and of excess electrons  in the oxide layer,
\bgea
n_e = \dst 2 \lb 2 \pi m_e k_b T /h^2  \rb^{3/2} \exp\lb -\frac{\mathbf{e} \phi}{k_b T}\rb ,
\\
n_i = \dst N_i \ \exp\lb -\frac{\mathbf{e} W_i}{k_b T}\rb ,
\enea
where $m_e$ is the mass of electron, $k_b$ Boltzman constant, $h$ Plank constant, $N_i$ is the concentration of sites available for hopping metal ions. Thus
$$
\bga
N =   \lb n_e n_i \rb^{1/2}=N_0 \lb \frac{T}{T_0} \rb ^{\frac{3}{4}}  \exp\lb -\frac{\mathbf{e} \lb \phi +W_i\rb }{2 k_b T}\rb ,
\\
N_0 = \lb 2 N_i \rb^{\frac{1}{2}} \lb 2 \pi m_e k_b T_0 /h^2  \rb^{3/4}
 \sim 1.5\  10^{27} \mbox{ m}^{-3} \mbox{ at }   T_0 \sim 750 K.
\ena
$$
The physical meaning of  $ W_i$  is the difference of chemical potentials for metal ions in the metal and the oxide; $\phi$ is the potential difference for electrons in the conduction bands of aluminum metal and the oxide (a semi-conductor); and the value of $V_0$ is determined from the condition that metal ion concentration at the interface with the oxide equals $n_i.$
In
Cabrera-Mott model, those ionization potentials may be considered as model parameters
\cite{mott}-\cite{11}.

Oxidation of metal occurs via tunneling of aluminum ions into the oxide layer, overcoming ionization potential of maximum height  $W >W_i$, and assisted by the self-consistent electric potential. Electric field  provides potential energy decrease for ion hopping from the bottom to the top of ionization potential, at a distance of $\al_0 \sim 0.4   \mbox{ nm}.$
This leads to an equation for the  metal-oxide interface radius $r_1:$ the normal velocity of the metal  boundary $u_n$ is determined by the electric field $(-\nabla V)$ at the boundary, appearing in a Gibbs factor,
\bgea
u_n \equiv \dst \frac{dr_1 }{dt} = - \Omega_1 \nu \ n_{2}
\exp\lb \frac{-e W}{k T }\rb
\exp\lb  \frac{q  e \al_0 \vert V^\prime (r_1) \vert }{k_b T }\rb;
\\
r_1 (0) = r_{10} .  
\label{MovBound}
\enea
Here  $r_{10} $   is the initial metal sphere  radius, which we assume to be  $22  \mbox{ nm},$  $\Omega_1 \approx 0.0166 \mbox{ nm}^3$ is the volume of oxide per aluminum ion,  $n_{2}\sim 10 \mbox{ nm}^{-2}$ is the number of metal ions per unit surface area, $\nu \sim 10^{12} s^{-1} $ is   frequency of tunneling attempts, $ q=3$ is aluminum valency.

In the Cabrera-Mott model, it is assumed that escaped metal ions migrate to the outer boundary of the oxide  where they react with the oxygen, while local Gibbs distribution of excess densities of electrons and metal ions inside the oxide is essentially unaffected. Due to the spherical symmetry, the radius of the oxide-oxidizer interface   $r_2$ changes  uniformly, and can be found from conservation of the number of metal ions, taking into account difference in volumes per metal ion in the metal and the oxide,
$  \lb r_2^3 -r_{20}^3  \rb = -  \kappa   \lb r_1^3 -r_{10}^3  \rb, $ $\kappa  \approx 0.386.$ Thus
\bgea
r_2 \equiv   r_2 \lb r_1 \rb= \lb r_{20}^3+ \kappa   \lb r_{10}^3 -r_1^3  \rb  \rb^\third , \kappa  \approx 0.386 .
\label{r2}
\enea
Here $r_{20}= r_{10} +\delta,$ where  $\delta  \sim 3 \mbox{ nm} $ is the initial oxide layer thickness.

For small metal particles, it is important to take {\it self-heating}  into account, due to heat released  by exothermic  aluminum oxidation, resulting in temperature increase   of the remaining metal and  oxide layer.
For nano-sized particles, temperature can be assumed to be uniform.
Thus
temperature can be computed based on reaction heat release and specific heats of reagents. Assuming constant specific heats,
\bgea
T \equiv T \lb r_1 \rb = T_0 + \dst\frac{\sigma H_{{\mbox{\fontsize{5pt}{11pt}\selectfont$Al$}}} 
   \rho_{{\mbox{\fontsize{5pt}{11pt}\selectfont$Al$}}}  \lb r_{10}^3 - r_1^3 \rb}{ c_{{\mbox{\fontsize{5pt}{11pt}\selectfont$Al$}}} \rho_{{\mbox{\fontsize{5pt}{11pt}\selectfont$Al$}}}r_1^3  + c_{{\mbox{\fontsize{5pt}{11pt}\selectfont$Al_2O_3$}}} \rho_{{\mbox{\fontsize{5pt}{11pt}\selectfont$Al_2O_3$}}} \lb  r_2^3 - r_1^3 \rb  }.
\label{T}
\enea
Here $\rho_{{\mbox{\fontsize{6pt}{11pt}\selectfont$Al$}}} $,  $\rho_{{\mbox{\fontsize{6pt}{11pt}\selectfont$Al_2O_3$}}} $ are densities of the  aluminum and oxide, $c_{{\mbox{\fontsize{6pt}{11pt}\selectfont$Al$}}} $,  $c_{{\mbox{\fontsize{6pt}{11pt}\selectfont$Al_2O_3$}}} $ are specifics heats per unit mass, $H_{{\mbox{\fontsize{5pt}{11pt}\selectfont$Al$}}}$ is the oxidation reaction enthalpy per unit mass of aluminum, and $\sigma$ is the proportion of released heat which is used up for self-heating.  We take reaction initiation temperature $T_0 =750 K. $ Since $r_2$ can be found from $r_1$ using (\ref{r2}), the   temperature $T$ in (\ref{T}) is determined by $r_1.$
Due to high enthalpy release in aluminum oxidation ($ \sim 24$ kJ per gram of aluminum ), adiabatic assumption $\sigma=1$ yields unrealistically  high maximum  temperature, even when  melting and vaporization heats are taken into account.
 We assume that   $\sigma=0.11 $ of the heat released
contributes to  self heating, while the rest is lost due to radiation, heat conduction, and convection. For such value of $\sigma$, the maximum reaction temperature, corresponding to $r_1=0, r_2= r_2(0)$  in (\ref{r2}),  (\ref{T}), will be $T_{M}\sim 2000 ^o C,$ which agrees with  experiments \cite{3}.

We use experimentally determined maximum temperature  to control the effect  of the  heat transfer process on oxidation rates.
Detailed modeling of the heat transfer is outside the scope of this Letter.
However, we note that an assumption of fairly large heat loss is sensible.  For   a nanoscale particle, the convection loss mechanism   is important. Indeed, in the steady state limit, heat loss of a small sphere due to convection can be modeled by the Newton's law of cooling, with the heat transfer coefficient $\eta,$   $\ls W m^{-2}K^{-1}\rs$,  given by $\eta =\frac{k}{r}$
\cite{16}, where $k,$ $\ls W m^{-1}K^{-1}\rs$,  is  air's thermal conductivity, and $r$ is Al sphere radius  (this limit is applicable for small Grashof numbers, which is the case for nano-spheres). Thus, the convective loss rate  $\eta (4 \pi r^2) \delta T$ is proportional to  $r$, while the total heat released due to chemical reaction is proportional to  $r^3$. As a result, convective heat loss becomes significant for small spheres. An estimate shows that for an aluminum  particle of radius 25 nm, all the heat released by oxidation  may be lost to steady convection in a time scale of $10^{-8}$ s,  which is comparable to the time scale of the final rapid phase of the oxidation process in our model.  Radiation loss, which is proportional to  $r^2 T^4$, contributes less than convection for the values of parameters in our model. 


We note that Eq. (\ref{MovBound}) for the metal boundary velocity  is similar to the model of metal sphere oxidation considered in  \cite{drez}, however we use  a different potential (a solution of the nonlinear Poisson equation (\ref{MottPot}) rather than the Coulomb potential),  and we also take the  self-heating effect into account. As a result, oxidation rates are dramatically increased. 


We have found the self-consistent  potential $V,$  solving numerically the boundary value problem for the Poisson equation (\ref{MottPot})  for various radii $r_1 $  (we have used the  Newton's method to solve a discrete version of the problem).  Our results for the potential are illustrated in Figure 1. Our values of ionization parameters  $V_0=0.65$ V, $\phi + W_i =1.5$ V, $W =1.7$ V,  $a=0.4$ nm are consistent with the data in the Cabrera-Mott paper \cite{mott}. We used $r_{10} = 22 $ nm for the initial metal radius, and $r_{20} = 25$ nm for the initial metal+oxide radius. When the remaining metal radius is  $r_1\leq r_{10}$, the radius of the metal +oxide $r_2$ can be found from Eq. (\ref{r2}), and the temperature, with the self heating taken into account, is determined by  (\ref{T}). We use  $T_0=750 K$ as the initial temperature.

We note that for the case of cylindrical symmetry,  the potential equation is equivalent to Painleve 3 (with particular values of parameters). 

Solution of the nonlinear Poisson equation  (\ref{MottPot})  enables us to compute the gradient of the potential  on the metal surface, and corresponding potential decrease due to electric field near metal-oxide interface, for various radii $r_1.$ In Figure 2, we show electric potential decrease at a distance $a=0.4 nm$ from the  metal-oxide interface  for the solution of nonlinear Poisson equation (solid  line), and compare it with a potential decrease computed using Coulomb potential (broken line). Potential decrease determines oxidation speed, via an exponential Gibbs factor in Eq. (\ref{MovBound}). Our results demonstrate that the nonlinear model for the potential, with the self-heating effect, yields considerably higher oxidation rates.

Using results for the gradient of the potential on metal-oxide boundary, we can find the radius of the metal sphere as a function of time, by solving the moving boundary  Eq. (\ref{MovBound}).  This equation is separable, since  it follows from Eqs. (\ref{r2}),(\ref{T})  that
the right hand side is a function of $r_1$ only. Thus,  a solution can be found by a numerical integration.
The  solution is shown in Figure 3. The overall oxidation time scale is in microseconds, much faster than for macroscopic particles. We further observe that in our model most of oxidation  occurs very quickly towards the end of oxidation process, since reaction rate dramatically increases with the temperature rise due to the self-heating.

Our model yields oxidation time of   32 $\mu s$ for a spherical aluminum nanoparticle of radius  $25$ nm at initial temperature of $750$ K. By using exactly the same ionization parameters and the self-heating mechanism, but taking the Coulomb potential rather than a solution of the self-consistent equation (\ref{MottPot}), oxidation time is calculated to be 2.5 ms, that is almost 100 times slower. As for the diffusion limit, it is known to correspond to the linear in the Coulomb potential $V^\prime$ approximation of the Eq.  (\ref{MovBound}). 
It is clear from Eq. (\ref{MovBound}) that the oxidation rate in  the linearized model will much slower than for the Mott model. Using the same model parameters and the self heating as in our model, oxidation time in the diffusion limit would be about 40 ms; and without the self-heating it is estimated to be 14 seconds at $T=750 K$. Experimentally, an explosive oxidation reaction of nanosized particles is initiated at a  temperature of about $750K$ (i.e below aluminum melting temperature),   with the reaction time scale estimated to be in microseconds \cite{3}. This comparison strongly suggests the Cabrera-Mott model  with a self-consistent potential as   the most likely reaction mechanism.

Figure 4 illustrates  the oxidation rate of change of mass of   aluminum   in a  metal-oxide  sphere with initial radius $25 \ nm$ , as a function of aluminum  mass and  temperature $T$.  Such a rate of change of reacting mass  appears e.g. in Semenov  explosion model, with $\frac{1}{m}\frac{dm}{dt}$ usually given by the Arrhenius factor $\exp (-\frac{E}{T} ) $ with a constant $E.$ However in Cabrera-Mott oxidation model, $E$ depends on the remaining mass and temperature (since electric potential drop in the self-consistent Cabrera-Mott model depends on the metal radius and temperature).

Nonlinearity in equation for the self-consistent electric potential,  together with the  self-heating effect, lead to significant increase of oxidation rate for nano-sized aluminum particles, compared to  the model with Coulomb potential and without self-heating.
Results of our modeling suggests oxidation time scales  in microseconds range for nano-sized aluminum particles, in agreement with the  experimental evidence \cite{3}. Results of computation of  oxidation time scale are very sensitive to values of ionization potentials, since they appear as Gibbs factors in the equation for boundary velocity. Therefore we expect that oxidation rate will be very sensitive to defects, and to deviation from spherical symmetry. 

 After this work was completed, we have learned that in a recent paper \cite{drez2} a different mechanism of oxidation initiation, based on high temperature run-off and the Coulomb potential, was proposed for a smaller, 10 nm radius, particle with very thin initial oxide layer. In our model with the nonlinear equation for the potential, rapid full oxidation of nanosized particle can occur at lower, experiment-matched, maximal temperature,  due to nonlinear effects for the potential near metal-oxidizer interface.  

We acknowledge  the financial support of this research by the National Science Foundation grant  0933140.


\section*{Figures}
\includegraphics[width= 3 in, height= 2.1in]{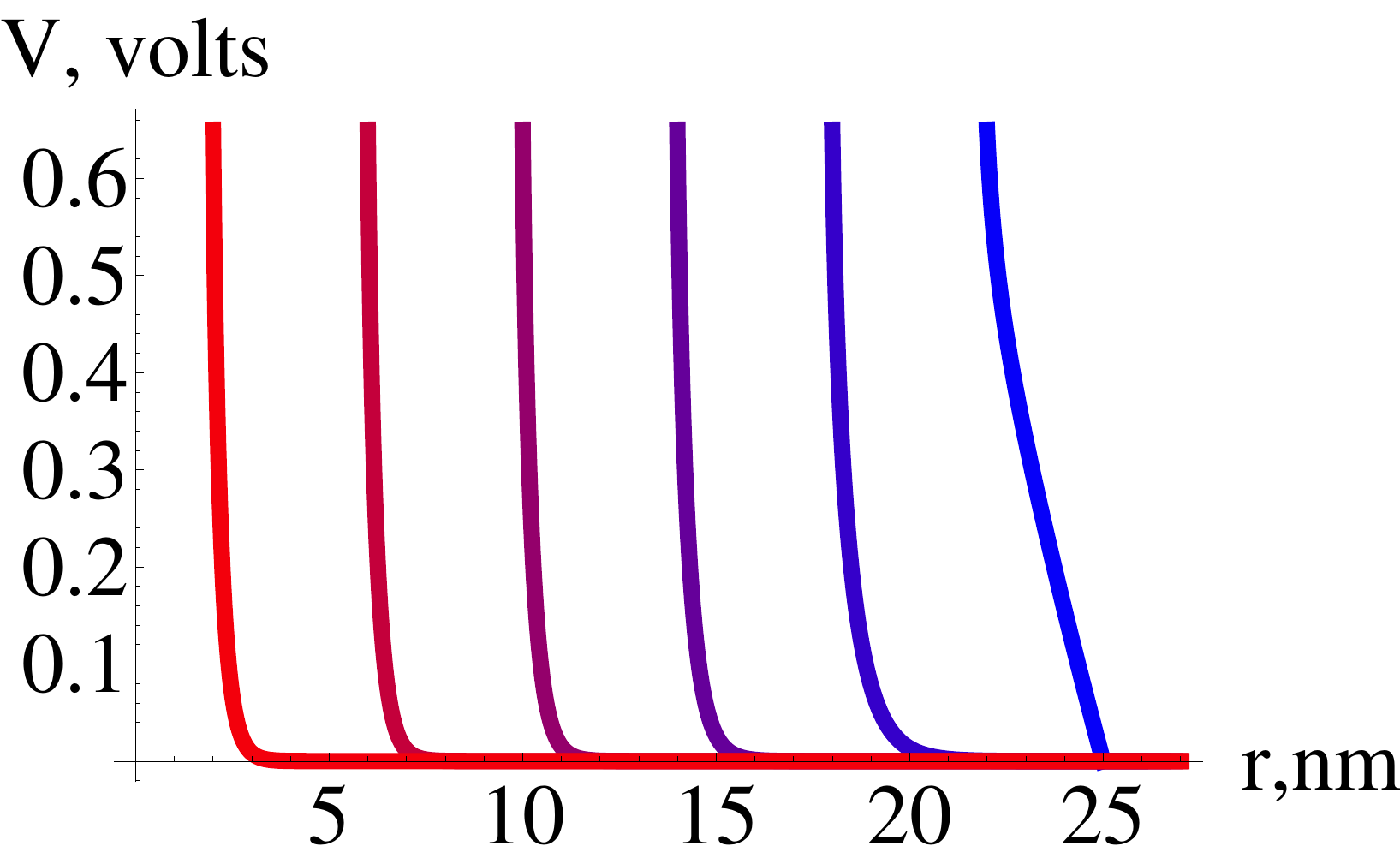}
\\
{\small {\bf Figure 1:} Self-consistent potential $V$ for $r_1 = 2, 6, 10, 14, 18, 22$ nm. Here $r_{10}= 22$ nm, $r_{20}= 25$ nm, $V_0 =0.65$ V,  $\phi + W_i =1.5 $ V ,  $W =1.7 $ V, $T_0 =750$ K.  }
\vspace{6mm}

\includegraphics[width= 3 in, height= 2.in]{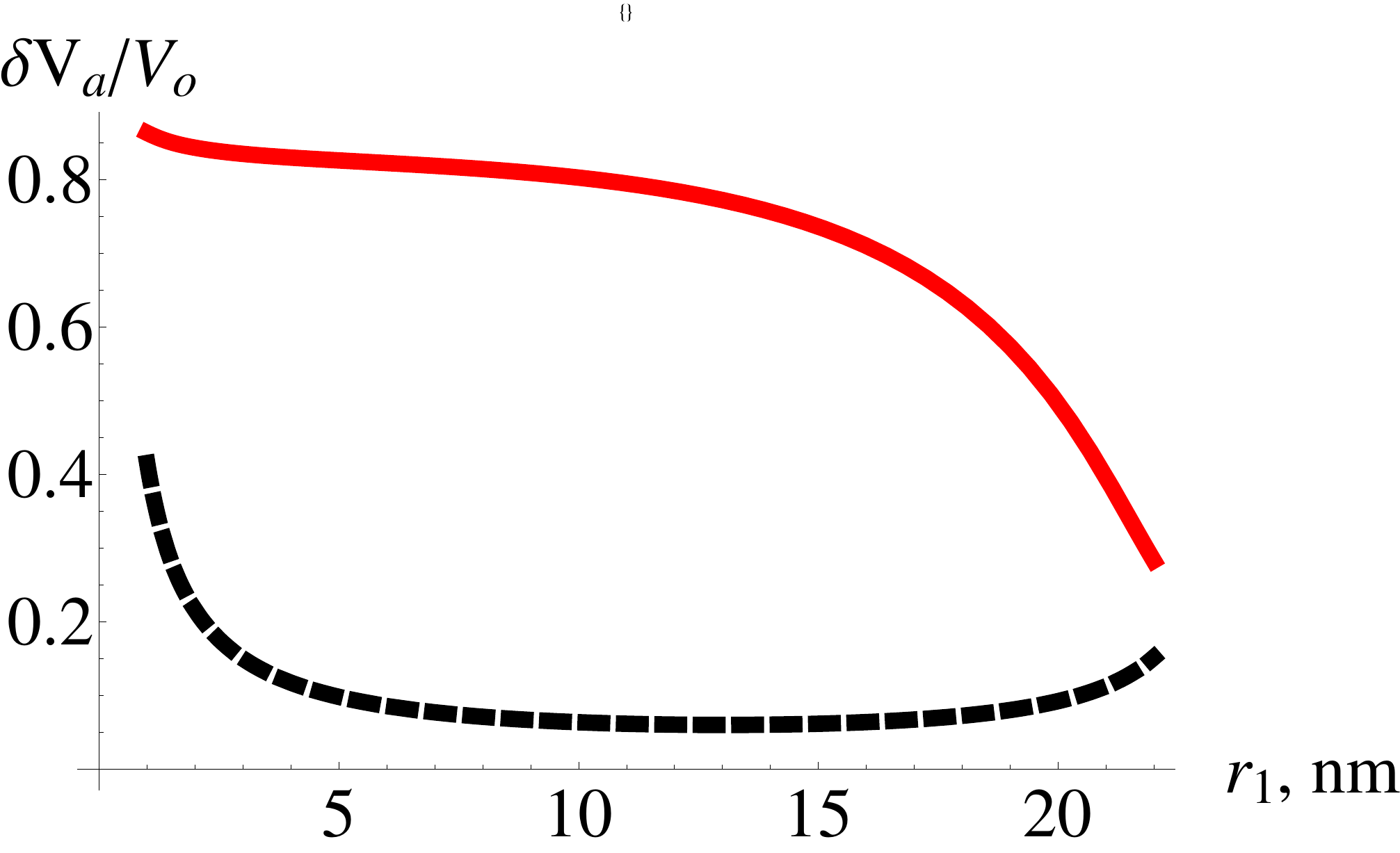}
\\
{\small {\bf Figure 2:} Potential drop  a distance $a =0.4$  nm  on the oxide side from aluminum metal-oxide interface, shown as a proportion of the total Cabrera-Mott potential $V_0=0.65 $ V, as a function of the remaining metal radius, and taking self-heating into account. Solid red curve: nonlinear self-consistent model for the potential; black dashed curve: Coulomb potential.  Here $r_{10}= 22$ nm, $r_{20}= 25$ nm,   $\phi + W_i =1.5 $ V ,  $W =1.7 $ V, $T_0 =750$ K , $T_{M} = 2273$ K. }

\newpage

\includegraphics[width= 5 in, height= 3.35 in]{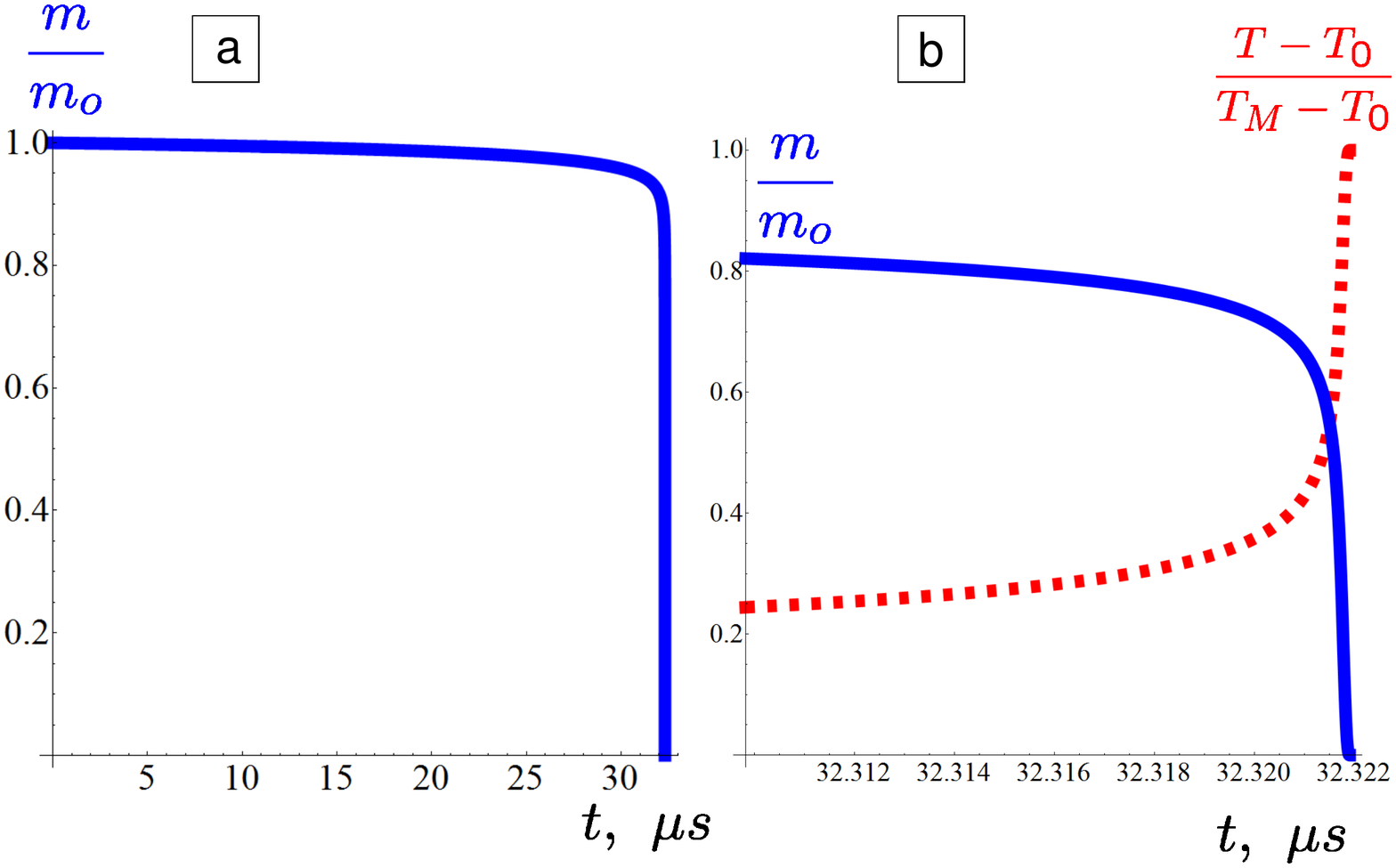}
\\
{\small {\bf Figure 3:} Oxidation time scales: (a) ratio of remaining aluminum  metal mass to initial aluminum mass, as a function of time. Here $r_{10}= 22$ nm, $r_{20}= 25$ nm,   $\phi + W_i =1.5 $ V ,  $W =1.7 $ V, $T_0 =750$ K , $T_{M} = 2273$ K,  $a =0.4$ nm. (b) final stages of oxidation shown; solid curve:  ratio of the remaining aluminum  metal mass to the initial aluminum mass, dashed curve: ratio of temperature increase to the maximal temperature increase}
\newpage

\includegraphics[width= 4.5 in, height= 3 in]{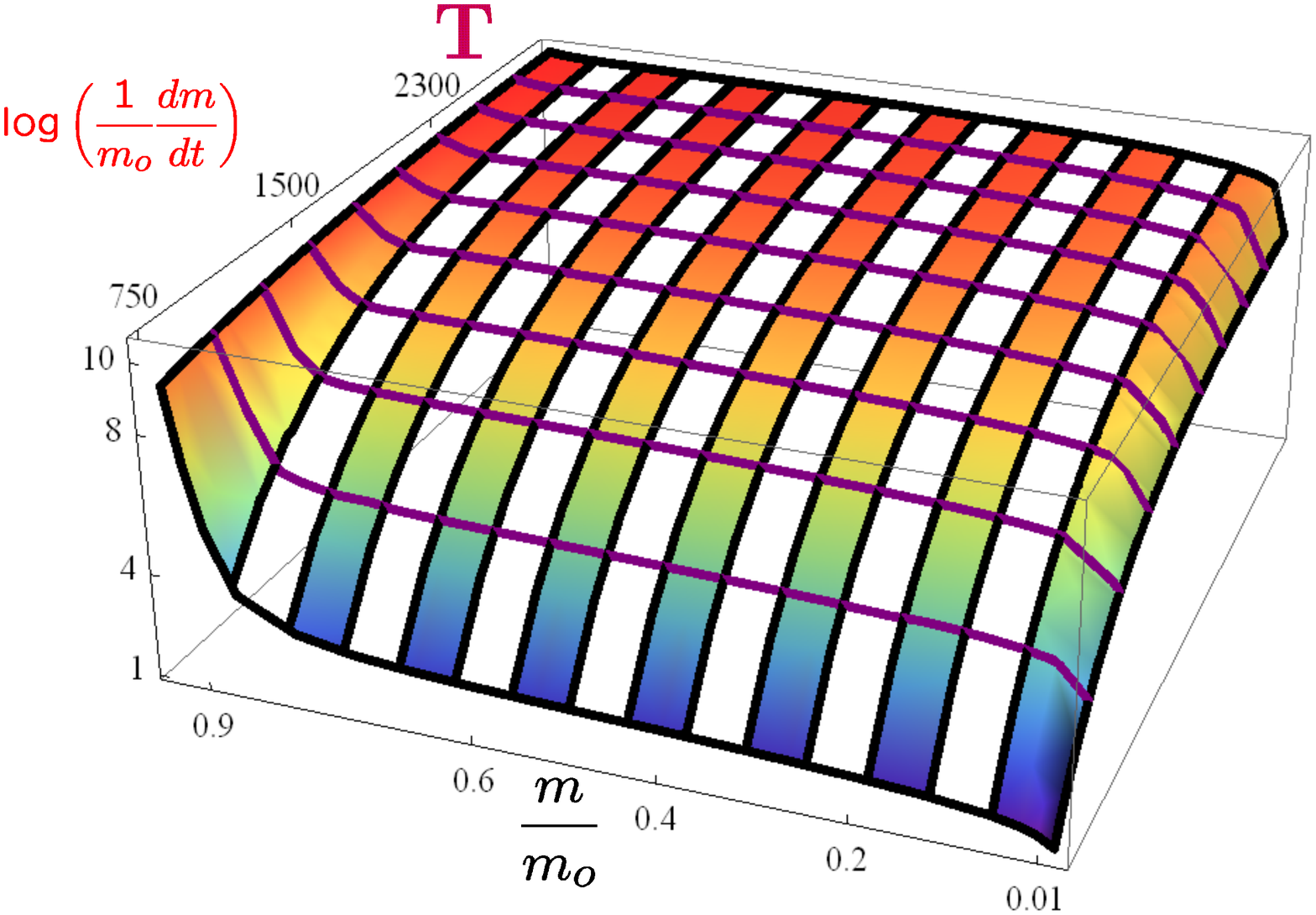}
\\
{\small {\bf Figure 4:} Oxidation rate of change of mass of metal aluminum  in a  metal-oxide  sphere of initial radius $25$ nm, as a function of metal mass $m$   and temperature $T$ (K), in $\log_{10}$ scale. Metal  mass is shown as a ration to the initial mass of the  particle $m_0.$ Here initial radius is 25 nm,   $\phi + W_i =1.5 $ V ,  $W =1.7 $ V, $T_0 =750$ K, $T_{M} = 2273$ K, $a =0.4$ nm. }

\end{doublespace}

\end{document}